\documentstyle[prl,aps,multicol,epsf,times]{revtex}

\newcommand{\om}{\omega}
\newcommand{\Li}{{\cal L}}

\begin{document}
\draft 
\title{A Stochastic Liouvillian Algorithm to Simulate
Dissipative Quantum Dynamics\\ With Arbitrary Precision}

\author{J\"urgen T. Stockburger and Chi H. Mak}

\address{Department of Chemistry, University of Southern California,
Los Angeles, CA 90089-0482, USA}

\maketitle
\begin{abstract}
An exact and efficient new method to simulate dynamics in 
dissipative quantum systems is presented. A stochastic Liouville 
equation, deduced from Feynman and Vernon's path-integral expression of 
the reduced density matrix, is used to describe the exact dynamics at 
any dissipative strength and for arbitrarily low temperatures.
The utility of the method is demonstrated by applications to a 
damped harmonic oscillator and a double-well system immersed in an
Ohmic bath at low temperatures.
\end{abstract}


\begin{multicols}{2} 
\narrowtext
Relaxation dynamics and fluctuations of a quantum system
in contact with a thermal bath are improtant for many different areas 
of chemistry and physics, such as reaction-rate 
theory\cite{hanggi,chandler}, 
ultrafast phenomena in chemistry and biochemistry\cite{marcus}, tunneling at
defects in solids\cite{Wipf} and quantum optics. While
some limiting cases permit Markovian or weak-coupling approximations to the
dissipation effects, the majority of others do not. 
Problems involving coherence\cite{Chakravarty-Leggett} are prime examples 
in which the Markovian approximation fails.

The exact treatment of dissipative dynamics presented here is based on
the path integral influence functional formalism\cite{Feynman-Vernon},
developed by Feynman and Vernon for interacting systems and
extensively used in dissipative quantum
mechanics\cite{Leggett,WeissBuch}. The main difficulty in evaluating
the complex-valued path integral involved is that the integrand is
{\em not} a local functional of the paths. Except for the case where
the memory effects decay on a short time scale\cite{nancy},
deterministic integration methods fail, making it prohibitively
expensive in both computation time and memory requirements. To date,
the only exact numerical approach to dissipative quantum dynamics with
arbitrary memory time has been the dynamical quantum Monte Carlo (QMC)
method. This method, however, is severely limited by the sign problem,
i.e. the signal-to-noise ratio of the numerical result tends to zero
exponentially with the time scale of the problem.

In this Letter, we deal with the problem with non-local influence
functionals arising from arbitrary long bath memory.
We recover fully time-local dynamics for the
ubiquitous case of Ohmic dissipation by applying a simple
Hubbard-Stratonovich transformation. 
The time-local path integral can then easily be solved by
propagating an equation of motion for the reduced density matrix. The
auxiliary quantity introduced by the transform appears in the dynamics
as an additional classical time-varying force. 
The functional integration over the
auxiliary quantity can be performed stochastically by interpreting it
as a {\em noise force}. This yields a new and elegantly simple algorithm:
Generate an ensemble of Gaussian noise trajectories with the
appropriate spectrum, propagate the system for each noise trajectory,
and average over the ensemble. In stark contrast to QMC methods, we
find a statistical error that is virtually independent of time.

At a microscopic level, dissipation is caused by the interaction of a
non-thermal system with a vast number of environmental modes at or
close to thermal equilibrium. Although cumulative effects of these
modes interacting with the system under study can have a drastic
effect on its dynamics, the coupling to each individual mode is
usually weak.  This justifies the paradigmatic model employed by Caldeira
and Leggett\cite{Caldeira-Leggett} to describe dissipative quantum 
systems -- a particle in an arbitrary potential interacting
linearly with a vast number of harmonic excitations,
\begin{equation}
H = {p^2\over 2 m} + V(q) + \sum_\nu {p_\nu^2\over 2 m_\nu}
+ {m_\nu \om_\nu^2\over 2}
 \left(x_\nu - {c_\nu\over m_\nu \om_\nu^2}q\right)^2 .
\end{equation}
The effect of the environmental modes is fully characterized by their
spectral density $J(\om) = {\pi\over 2} \sum_\nu {c_\nu^2\over
m_\nu\om_\nu} \delta(\om - \om_\nu)$, which takes on the form $J(\om)
= \eta\omega$ in the Ohmic case, $\eta$ being the classical friction
constant.

The time-dependent reduced density matrix is then given by a double
path integral
\begin{equation} \label{rho_PI}
\rho(q_{\rm f},q'_{\rm f};t) =
\int_{q_{\rm i}}^{q_{\rm f}} \!\!\!{\cal D}[q]
\int_{q_{\rm i}}^{q'_{\rm f}} \!\!\!{\cal D}[q']
e^{{i\over\hbar}( S_0[q] - S_0[q'])} F[q,q']. \label{foo}
\end{equation}
Here $S_0[q]$ is the action of the undamped quantum system,
and its interactions with the environment are  
incorporated into a complex-valued influence functional $F[q,q'] =
\exp(-\Phi'/\hbar - i\Phi''/\hbar)$ with 
\begin{eqnarray}
\Phi'[q,q'] &=& \int_{t_0}^t \!\!\!dt' 
\int_{t_0}^{t'} \!\!\!dt''
(q(t')-q'(t')) \nonumber\\
&& \times L'(t'-t'') (q(t'')-q'(t'')), \\ \label{phipp}
\Phi''[q,q'] &=& {1\over 2} \eta \int_{t_0}^t \!\!\!dt'
(q(t')-q'(t')) \nonumber\\
&& \times (\dot q(t')+\dot q'(t')). \label{bar}
\end{eqnarray}
$L'(t)$ is the real part of the autocorrelation function of the
collective bath coordinate $\sum_\nu c_\nu x_\nu$,
averaged over an ensemble of free oscillators.

This influence functional describes a `factorized' initial
preparation, i.e. with the particle constrained to an initial position
$q_i$ for times $t<t_0$ \cite{inifoot} with the environment fully
relaxed. Equilibrium correlation functions may be calculated by
pushing the preparation back to a sufficiently large negative time
$t_0<0$ \cite{WeissBuch} and inserting measurement operators at times
$0$ and $t$.

In the following we outline the numerical solution of these equations
using a novel technique we call the SLED (Stochastic Liouville
Equation for Dissipation) method. SLED is 
based on the same principles as another method we recently introduced 
-- the CSQD (Chromostochastic Quantum Dynamics) method\cite{csqd}.
Whereas CSQD was developed to treat discrete systems, 
SLED solves extended systems using a stochastic Liouvillian formalism.

The primary obstacle in trying to translate eqs. (\ref{foo}) --
(\ref{bar}) into a recursion relation or an equation of motion for
$\rho$ lies in the interaction kernel $L'(t)$. Its range is often
large, becoming divergent for $T\to 0$. This problem of retarded
self-interactions mediated by $L'(t)$ can be solved, albeit at the
cost of introducing an additional path variable.  The exponential of
the non-local action $\Phi'[q,q']$ can be decomposed into time-local
phase factors,
\begin{eqnarray} \label{noise}
&&\exp(-\Phi'[q,q']) = \nonumber \\
&&\int {\cal D}[\xi] \, W[\xi] 
\exp\left\{ -i 
\int_{t_0}^t \!\!\!dt' \xi(t')(q(t')-q'(t')) \right\}.
\end{eqnarray}
The distribution function $W[\xi]$ is real and Gaussian, with
$\langle\xi(t)\xi(t')\rangle_W = L'(t-t')$, and normalizable through
the condition $\Phi'[q,q] \equiv 0$. Formally, this decomposition is just a 
Hubbard-Stratonovich transformation in a function space over the
interval $[t_0,t]$. Equation (\ref{noise}) is also equivalent to the
construction of an influence functional for a classical colored noise
source\cite{Feynman-Vernon}, and as such, we will interpret the
function $\xi(t)$ as a noise trajectory and the measure ${\cal
D}[\xi] W[\xi]$ as the probability measure of an ensemble of noise
trajectories. For each noise trajectory, the reduced density matrix is
then propagated deterministically according to the equation of motion 
\begin{equation}
i\hbar\dot \rho = \Li_0 \rho + \Li_\eta \rho + \xi [q,\rho] \label{eom1}
\end{equation}
where $\Li_0$ is the Liouvillian of the undamped system, and the friction
term $\Phi''$ gives rise to an additional operator in the Liouvillian 
\begin{equation}
\Li_\eta = -{\eta\over 2 m}[q,\{p,\cdot\}] .
\end{equation}

The spectrum of the noise is determined by the spectral density and
the temperature. In the Ohmic case, it is given by
$ S(\om)
=
\eta\om\coth(\hbar\beta\om/ 2)$
up to an arbitrary but necessary cutoff frequency $\om_c$.
For computational purposes, it is advantageous to split off a
`classical' white noise part $S_{\rm cl}(\om) = 2\eta/\hbar\beta$ from this
spectrum because it can be treated by a damping
term
\begin{equation}
\Li_d = - i{\eta\over\hbar \beta} [q,[q,\cdot]]
\end{equation}
that can be incorporated into the deterministic equation of motion 
in the Liouvillian rather than explicitly in the numerically generated 
noise field\cite{Leggett-liou}. The full equation of motion then reads
\begin{equation}
i\hbar\dot \rho = \Li_0 \rho + \Li_\eta \rho + \Li_d \rho
 + \tilde\xi [q,\rho] \label{eom2}
\end{equation}
where the noise spectrum of $\tilde\xi$ is $\tilde S(\om) = S(\om) -
S_{\rm cl}(\om)$. It has been shown previously \cite{Leggett-liou}
that using eq. (\ref{eom2}) without the noise force $\tilde\xi$ is a
limiting case that becomes exact only if $S(\om) = S_{\rm cl}(\om)$,
i.e., for very high temperatures $kT \gg \hbar\om_c$. Comparing 
the exact dynamics expressed in eq.~(\ref{eom2}) to
this limiting case, we find a remarkable interpretation of our
formalism: When a linear environment is treated classically, one
generally incurs an error due to the missing quantum fluctuations. But
by explicitly substituting an external noise force for the quantum
fluctuations one actually regains the {\em exact} quantum dynamics at
{\em any} temperature!

In the following we give examples, solving eq. (\ref{eom2}) in the
position representation using the split-propagator technique. The
alternating direction implicit method is used for the operator $\Li_0
+ \xi[q,\cdot]$ and an explicit method for the remaining term
$\Li_\eta + \Li_d$. This choice is appropriate as it preserves the
trace\linebreak of $\rho$.

Our first test consists in the computation of the equilibrium
correlation function $S(t) = {1\over 2} \langle q(t)q(0) +
q(0)q(t)\rangle$ of the damped harmonic oscillator, shown in Fig. 1,
and its response function $\chi(t) =
{i\over\hbar}\langle[q(t),q(0)]\rangle$. At relatively high
temperature, $kT = 2\hbar\om_0$, this calculation converges with only
a handful trajectories and reproduces an almost classical result, as
expected in this temperature regime. The initial value $S(0)$
corresponds exactly to the thermal expectation value $\langle
q^2\rangle = kT/m\om_0^2$. At low temperature, $kT = 0.2\hbar\om_0$,
the fluctuation amplitude is no longer temperature-dependent, but
is instead indicative of zero-point fluctuations. 
(Note that the initial amplitude
for the damped oscillator is slightly reduced from the undamped value
$\langle q^2\rangle = \hbar/2m\om_0$\cite{Weiss_remark}).  In the same
test, the susceptibility $\chi(t)$ of the quantum harmonic oscillator
is determined. We find excellent agreement with the known analytic
result $\chi(t) \equiv \chi_{\rm cl}(t)$. The high-temperature
calculation completes within minutes, while the low-temperature
calculation takes about $10^3$ trajectories for a very small absolute
error $<0.007$, or 40 CPU hours on a midrange IBM Risc processor using
a preliminary, non-optimized code.

For a more elaborate example, we use the quartic double-well potential
$V(q) = q^4/4 - q^2$, where natural units for time, mass, and energy
are implied. In the undamped case, this system shows complex
dynamics including both tunneling and vibrational oscillations. Fig. 2
shows the equilibrium dynamics of this system for different values of
friction and temperature. For high friction, $\eta = 0.5$, the system
shows a monotonous decay, characteristic of incoherent relaxation
(solid line). When the friction is lowered to $\eta = 0.05$, a
qualitative change to complex, anharmonic oscillatory dynamics is
observed (dotted line). For the same friction, lowering the
temperature leads to another qualitative change in the dynamics, resulting in
damped oscillations with two manifestly distinct frequency scales
(dashed line). The low-frequency component represents coherent
tunneling, with a frequency related to the splitting between the two
lowest energy eigenstates of the double well, and the high-frequency
part derives from higher excited states.

In conclusion, we have developed a novel algorithm for the dynamics of
extended dissipative quantum systems, which is highly efficient. It is
also shown to be accurate on first principles and by explicit numerical 
tests. We have presented simulation data both for the damped harmonic
oscillator and the dissipative double-well system which show the
practical feasibility and power of this method. 
Empirically, we find that the statistical error of our results
is extremely small,
asymptotically approaching a {\em constant} value for long times. 
This implies a CPU requirement that scales {\em linearly} with
the system time $t$, compared to the {\em
exponential} rise for conventional quantum Monte Carlo methods.
The SLED algorithm is also easily implemented on parallel
computers.  The system propagation for different noise realizations can
be performed in parallel without any dependencies beyond the
accumulation of final results.

We are currently extending the SLED method to study dynamics on
electronically coupled energy surfaces in condensed-matter
problems. The problem of greatest interest is the dynamics of electron
transfer reactions in the inverted region.  To model a bath with
an arbitrary cutoff frequency, the SLED method has to be extended to
treat systems with two electronic states coupled to a solvent mode
which is in turn coupled to an Ohmic bath\cite{garg}.

This research has been supported by the National Science Foundation
under grant CHE-9528121.
CHM is a NSF Young Investigator (CHE-9257094),
a Camille and Henry Dreyfus Foundation Camille Teacher-Scholar and a
Alfred P. Sloan Foundation Fellow.
Computational resources have been
provided by the IBM Corporation under the SUR Program at USC.

\begin{figure}
\epsfxsize=0.95\columnwidth
\centerline{\epsffile{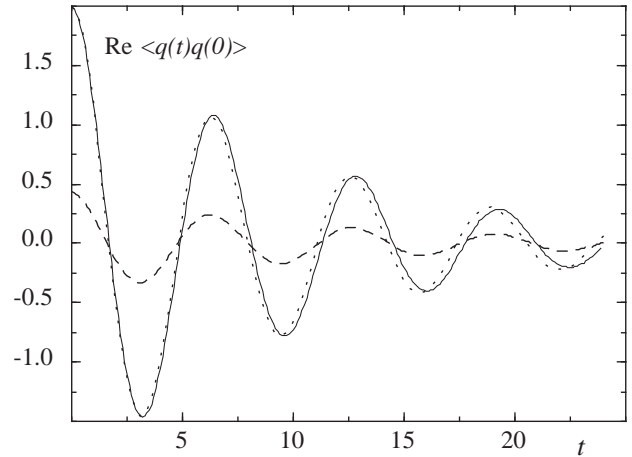}}
\caption[]{Equilibrium Fluctuations of the damped harmonic oscillator
at $T=2$ (solid line) and $T=0.2$ (long dash) with $\hbar=k=m=1$,
$\om_0=1$, $\eta = 0.2$. Short dashes indicate the analytic
high-temparature solution\cite{Weiss_remark} for $T=2$.}
\end{figure}

\begin{figure}
\epsfxsize=0.95\columnwidth
\centerline{\epsffile{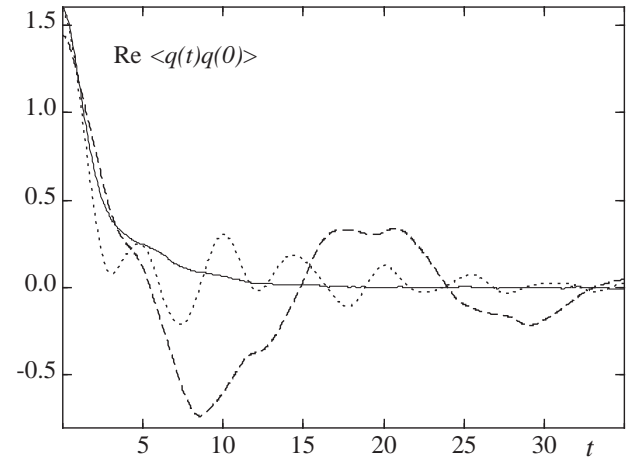}}
\caption[]{Equilibrium fluctuations of a dissipative double-well
system for different friction constants. Solid line -- dynamics showing
incoherent decay for $\eta=0.5$, $T=1$. Dotted line -- dynamics showing
predominantly anharmonic vibronic oscillations for $\eta= 0.05$, $T=1$. 
Dashed line -- dynamics showing predominantly tunneling oscillations for 
$\eta=0.05$, $T=0.5$.}
\end{figure}

\end{multicols}

\begin{references}
\bibitem{inifoot} The reduced density matrix at $t_0$ need not be
localized at $q_i$. The thermal density matrix of the undamped
system is often a more convenient choice for the initial condition.
\end{references}
\end{document}